\documentclass[10pt]{article}
\usepackage[OE]{express}

\begin{document}

\title{Conversion of entangled states with nitrogen-vacancy centers coupled to microtoroidal resonators}

\author{Y. Q. Ji,\authormark{1,2} X. Q. Shao,\authormark{1,2} and X. X. Yi\authormark{1,2,*}}

\address{\authormark{1}Center for Quantum Sciences and School of Physics,
Northeast Normal University, Changchun 130024, China\\
\authormark{2}Center for Advanced Optoelectronic Functional Materials Research, and Key Laboratory for UV Light-Emitting Materials and Technology of Ministry of Education, Northeast Normal University, Changchun 130024, China}

\email{\authormark{*}yixx@nenu.edu.cn} 

\begin{abstract}
We propose efficient schemes for converting three-photon, four-photon and five-photon GHZ state to a $W$ state or Dicke state, respectively with the nitrogen-vacancy (N-V) centers via single-photon input-output process and cross-Kerr nonlinearities. The total success probability can be improved by iterating the conversion process for the case of three-photon and five-photon while it does not require iteration for converting four-photon GHZ state to a $W$ state. The analysis of feasibility shows that our scheme is feasible for current experimental technology.
\end{abstract}



\section{Introduction}
It is well known that quantum entanglement is the basic resource in quantum information processing (QIP). It is widely used in quantum teleportation~\cite{001,002,003}, quantum key distribution~\cite{004,005,006}, quantum secret sharing~\cite{007,008,009,010} and quantum secure direct communication~\cite{011,012,013,014,015,015x,015y}. Furthermore, it is even considered as an important effect in living biological bodies in recent years, for instance, the entanglement may be related to Avian compass~\cite{0151}, and the entanglement and teleportation using living cell is also possible~\cite{0152}. Owing to its importance, many theoretical and experimental efforts for generating entanglement have been one focus of the current study~\cite{0153,S1,0154,0155,0156,S2,0157,0158,0159}. There are two most common classes of entangled states for a multipartite system, i.e., GHZ state and $W$ state, which cannot be converted to each other by local operations and classical communication (LOCC)~\cite{016,017}. For the GHZ state, the entanglement completely degrades with the loss of any one of the qubits. However, the $W$ state is robust against the loss of one qubit, when one qubit is discarded, the remaining qubits can be still entangled with each other. Due to this two kinds of entangled states can be used for performing different tasks in QIP~\cite{018,019,020,021,022,023,024,025}, naturally, the question arises: how the two kinds of entanglement can be converted into each other?

Recently, many research have been focused on the conversion between GHZ state and $W$ state~\cite{026,027,028,029,030,031}. Walther \emph{et al}. described a specific method based on local positive operator valued measures and classical communication to convert a $N$-qubit GHZ state into an approximate $W$ state experimentally and implemented this scheme in the 3-qubit case in 2005~\cite{026}. Subsequently, Tashima \emph{et al}. proposed and experimentally demonstrated a transformation of two EPR photon pairs distributed among three parties into a three-photon $W$ state using LOCC in 2009~\cite{027}. More recently, Wang \emph{et al}. proposed a linear-optics-based scheme for local conversion of four EPR photon pairs distributed among five parties into four-photon polarization-entangled decoherence-free states~\cite{028}. Song \emph{et al}. proposed a scheme for converting a $W$ state into a GHZ state with a deterministic probability through a dissipative dynamics process~\cite{029}. These works provides a new way to implement the conversion of entangled states.

As a promising candidate for the quantum computer, nitrogen-vacancy (N-V) centers in diamond are coupled to the microresonator with a quantized whispering-gallery mode (WGM)~\cite{032,033,034,035} shows great advantage and practical feasibility as they have a long lifetime even at room temperature and can be manipulated and detected by optical field
or microwave~\cite{036,037}. Hence, the N-V center, consists of nearest-neighbor pair of a nitrogen atom substituted for a carbon atom and a lattice vacancy in diamond, becomes one of the most potential carrier of quantum information due to these characteristics.

Motivated by the former works, we suggest a high-fidelity CNOT gate between two photons can be constructed resorting to N-V center through microtoroidal resonator (MTR) assisted single-photon input-output process. A high fidelity can be achieved. And the schemes of entangled states conversion are presented for converting three-photon, four-photon and five-photon GHZ state to a $W$ state or Dicke state by applying the CNOT gate and cross-Kerr nonlinearities. The schemes has the following advantages: the minimum success probability of conversion a three-photon GHZ state to a $W$ state is 3/4, while by iterating the conversion process,
the total success probability can be to unit, but the conversion for four-photon GHZ to a $W$ state does not need iteration. For five-photon GHZ state, the total success probability for converting to a $W$ state can be to 1/3 by iterating the conversion process. Meanwhile, we can obtain a Dicke state $|D_{5}^{(2)}\rangle$ with the success probability 2/3.

\section{Physical model}
\begin{figure}
\centering\scalebox{0.5}{\includegraphics{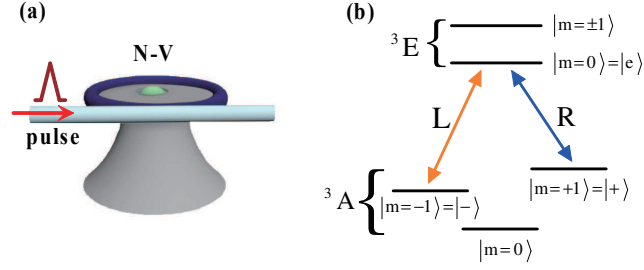}}
\caption{\label{P1} Diagrammatic illustration of basic model. (a) An N-V center is confined to a MTR and a single photon pulse is introduced to interact with the N-V center. (b) The electron energy level configuration of the N-V center and relevant transition coupling with corresponding polarization photon.}
\end{figure}
At the beginning, we provide a brief background on the basic unit of the N-V centers and microcavity system. In order to facilitate manipulating and extending the MTR-N-V center system, the N-V center should be close to the surface of diamond and the fibre transmitting input pulse is close to the resonator~\cite{038}. As illustrated in Figs.~\ref{P1}(a), an N-V center is fixed on the surface of a MTR and can be coupled to the cavity mode. And the electron energy level configuration of N-V center is shown in Figs.~\ref{P1}(b). The zero-field splitting and the Zeeman splitting from an external magnetic field determine the ground levels $|^{3}A, m_{s}=0\rangle$, $|-\rangle$ and $|+\rangle$~\cite{039,040}.
The information is encoded on the ground states $|-\rangle$ and $|+\rangle$. The excited state $|A_{2}\rangle$ acts as auxiliary level $|e\rangle$~\cite{040,041}. The level transition between $|-\rangle$ ($|+\rangle$) and $|e\rangle$ is resonantly coupled to the left (right) circularly polarized photon $|L\rangle$ ($|R\rangle$) in MTR.

Consider a single-photon pulse with frequency $\omega_{p}$ input in a MTR cavity with the mode frequency $\omega_{c}$. Under the assumption of weak excitation limit, i.e., $\langle \sigma_{z}\rangle=1$. So, the Langevin equations of motion are solvable for the lowering operators of the cavity and the N-V center. According to the cavity input-output process~\cite{042,0421,0422} and adiabatical elimination of the cavity mode leads to the reflection coefficient as~\cite{042}
\begin{eqnarray}\label{01}
r(\omega_{p})=\frac{\left[i(\omega_{c}-\omega_{p})-\frac{\kappa}{2}\right]\left[i(\omega_{0}-\omega_{p})+\frac{\gamma}{2}\right]+g^{2}}
{\left[i(\omega_{c}-\omega_{p})+\frac{\kappa}{2}\right]\left[i(\omega_{0}-\omega_{p})+\frac{\gamma}{2}\right]+g^{2}}
\end{eqnarray}
where $\omega_{0}$ is the transition frequency between energy levels $|-\rangle$ and $|e\rangle$. $\kappa$ and $\gamma$ are the cavity damping rate and the N-V center dipolar decay rate, respectively, and $g$ is the coupling strength of the cavity to the N-V center.
If the N-V centers are uncoupled from the cavity, i.e. $g=0$, Eq.~(\ref{01}) changes to
\begin{eqnarray}\label{02}
r_{0}(\omega_{p})=\frac{i(\omega_{c}-\omega_{p})-\frac{\kappa}{2}}{i(\omega_{c}-\omega_{p})+\frac{\kappa}{2}}
\end{eqnarray}
If a single polarized photon $|L\rangle$ is reflected from a MTR cavity after running into an N-V center prepared in $|-\rangle$ or $|+\rangle$,
it will experience a phase shift $e^{i\phi}$ or $e^{i\phi_{0}}$ from Eq.~(\ref{01}) or (\ref{02}) owing to optical Faraday rotation, respectively.
While if a single $|R\rangle$ polarized photon is reflected, it will experience a phase shift $e^{i\phi_{0}}$ regardless of what the electronic
spin states are. Choosing the resonant condition $\omega_{c}=\omega_{0}=\omega_{p}$ and $g^{2}\geq25\kappa\gamma$, then
\begin{eqnarray}\label{03}
r(\omega_{p})\simeq1,~~~~r_{0}(\omega_{p})=-1.
\end{eqnarray}
In conclusion, after a $\pi$ phase shifter on the photon output path, the systemic state between the photon and the N-V centers can
be expressed as
\begin{eqnarray}\label{04}
&&|R\rangle|+\rangle\rightarrow|R\rangle|+\rangle,~~~~|R\rangle|-\rangle\rightarrow|R\rangle|-\rangle,  \cr
&&|L\rangle|+\rangle\rightarrow|R\rangle|+\rangle,~~~~|L\rangle|-\rangle\rightarrow-|L\rangle|-\rangle.
\end{eqnarray}
In the following, we investigate how to construct a deterministic CNOT gate and implement conversion of entangled states between photonic qubits.

\section{Implementing of deterministic CNOT gate for photonic qubits}
\begin{figure}
\centering\scalebox{0.5}{\includegraphics{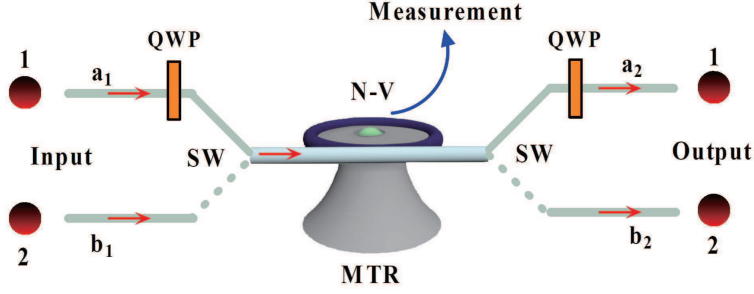}}
\caption{\label{P2}Schematic of deterministic photonic CNOT gate with an N-V center confined to a MTR. QWP is a quarter-wave plate which performs the transformations $|R\rangle\rightarrow\frac{1}{\sqrt{2}}(|R\rangle+|L\rangle)$, $|L\rangle\rightarrow\frac{1}{\sqrt{2}}(|R\rangle-|L\rangle)$, and SW is optical switch.}
\end{figure}
Consider two photons initially prepared in an arbitrary state $\psi_{p}=\alpha|RR\rangle+\beta|RL\rangle+\gamma|LR\rangle+\delta|LL\rangle$,
where $|\alpha|^{2}+|\beta|^{2}+|\delta|^{2}+|\gamma|^{2}=1$, and N-V center in the state $\psi_{N}=\frac{1}{\sqrt{2}}(|+\rangle+|-\rangle)$. Photon 2 acts as control qubit and photon 1 acts as target qubit. The schematic is shown in Fig.~\ref{P2}. The two photons 1 and 2 come in succession to the MTR. The action of the quarter-wave plate (QWP) is given by transformation $|R\rangle\rightarrow\frac{1}{\sqrt{2}}(|R\rangle+|L\rangle)$
and $|L\rangle\rightarrow\frac{1}{\sqrt{2}}(|R\rangle-|L\rangle)$. The optical switch (SW) controls the photon 1 passing through the MTR firstly
and then the photon 2. c-PBS is circular polarization beam splitter which transmits the right-circular polarization photon $|R\rangle$ and
reflects the left-circular polarization photon $|L\rangle$. In Fig.~\ref{P2}, before the photon 2 enters into the MTR, a Hadamard gate
operation, which can be achieved by a $\pi/2$ microwave pulse, is performed on N-V center to accomplish the transformation $|+\rangle\rightarrow\frac{1}{\sqrt{2}}(|+\rangle+|-\rangle)$
and $|-\rangle\rightarrow\frac{1}{\sqrt{2}}(|+\rangle-|-\rangle)$. After the photon 2 passing through the MTR, the N-V center is rotated by
the Hadamard gate transformation again. Finally, the total state of two photons with one N-V center is transformed into
\begin{eqnarray}\label{05}
\psi_{p}\otimes\psi_{N}&\rightarrow&\frac{1}{\sqrt{2}}(\alpha|RR\rangle+\beta|LL\rangle+\gamma|LR\rangle+\delta|RL\rangle)_{1,2}\otimes|+\rangle \cr
&&+\frac{1}{\sqrt{2}}(\alpha|LR\rangle+\beta|RL\rangle+\gamma|RR\rangle+\delta|LL\rangle)_{1,2}\otimes|-\rangle.
\end{eqnarray}
After the measurement performed on the N-V center, the CNOT gate between photons 1 and 2, which is unequivocally associated to
the measurement results of the N-V center in the $|+\rangle$, $|-\rangle$ basis, is achieved (see Table~\ref{tab1}).
\begin{table}
\begin{center}
\caption{\label{tab1} The measurement results and corresponding single-qubit operations on photons 1 and 2 in the case of the CNOT gate.}
\begin{tabular}{c c c}
  \hline
  ~~~~ N-V center ~~~~&~~~~ Photons 1 and 2 ~~~~&~~~~ Operations ~~~~ \\
  \hline
  $|+\rangle$ & $\alpha|RR\rangle+\beta|LL\rangle+\gamma|LR\rangle+\delta|RL\rangle$ & $I^{1}\otimes I^{2}$ \\
  $|-\rangle$ & $\alpha|LR\rangle+\beta|RL\rangle+\gamma|RR\rangle+\delta|LL\rangle$ & $\sigma_{x}^{1}\otimes I^{2}$ \\
  \hline
\end{tabular}
\end{center}
\end{table}

\section{Conversion of entangled states}
In this section, we illustrate how to convert a GHZ state to a W state in a deterministic
way based on single-photon input-output process and cross-Kerr nonlinearity.
\subsection{Conversion of a three-photon GHZ state to a three-photon $W$ state}
\begin{figure}
\centering\scalebox{0.45}{\includegraphics{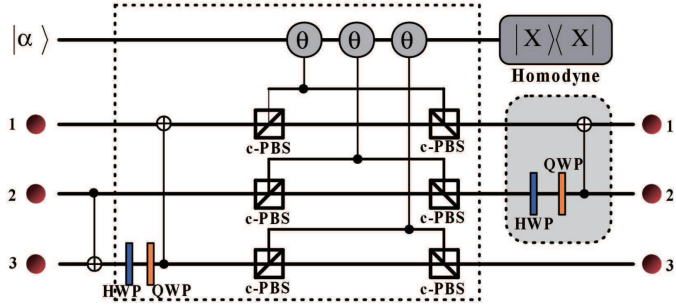}}
\caption{\label{P3}Schematic setup for converting a three-photon GHZ state to a three-photon $W$ state. Here HWP denotes a half-wave plate
which performs the transformations $|R\rangle\rightarrow|L\rangle$ and $|L\rangle\rightarrow|R\rangle$. QWP denotes a quarter-wave
plate which performs the transformations $|R\rangle\rightarrow\frac{1}{\sqrt{2}}(|R\rangle+|L\rangle)$ and $|L\rangle\rightarrow\frac{1}{\sqrt{2}}(|R\rangle-|L\rangle)$. c-PBS denotes a polarization beam splitter in the circular
basis that transmits $R$ photons and reflects $L$ photons. The action of the cross-Kerr nonlinearity puts a phase shift $\theta$ on
the probe beam only if a photon was present in that mode.}
\end{figure}
Assume that photons 1, 2 and 3, as shown in Fig.~\ref{P3}, are in the following state
\begin{eqnarray}\label{06}
|\psi\rangle_{0}=\frac{1}{\sqrt{2}}(|RLR\rangle+|LRL\rangle)|\alpha\rangle,
\end{eqnarray}
then the photons firstly pass through a CNOT gate, where photon 2 act as the control qubit and photon 3 act as the target qubit. Then the
photon 3 passes through a HWP, whose action is to make $|R\rangle\rightarrow|L\rangle$ and $|L\rangle\rightarrow|R\rangle$, and
a QWP, performing the transformations $|R\rangle\rightarrow\frac{1}{\sqrt{2}}(|R\rangle+|L\rangle)$ and $|L\rangle\rightarrow\frac{1}{\sqrt{2}}(|R\rangle-|L\rangle)$ in sequence. At last, the photons pass through a CNOT gate, with photon 3 being
control qubit and photon 1 being target qubit. The resulting state of the whole system is given by
\begin{eqnarray}\label{07}
|\psi\rangle_{1}=\frac{1}{2}(|RLR\rangle+|LRR\rangle+|RRL\rangle+|LLL\rangle)|\alpha\rangle.
\end{eqnarray}
After the three-photon state and a coherent probe beam couple with a cross-Kerr nonlinearity medium, the evolution
of the whole system can be described as
\begin{eqnarray}\label{08}
|\psi\rangle_{2}=\frac{1}{2}(|RLR\rangle+|LRR\rangle+|RRL\rangle)|\alpha e^{i\theta}\rangle+\frac{1}{2}|LLL\rangle|\alpha e^{i3\theta}\rangle,
\end{eqnarray}

For cross-Kerr nonlinearity, the Hamiltonian can be described as $H=\hbar\chi a_{s}^{\dag}a_{s}a_{p}^{\dag}a_{p}$~\cite{043,044,045},
here $a_{s}^{\dag}$ and $a_{p}^{\dag}$ are the creation operations, $a_{s}$ and $a_{p}$ are the destruction operations. $\chi$ is the
coupling strength of the nonlinearity. For a signal state $|\phi\rangle_{s}=\alpha|0\rangle_{s}+\beta|1\rangle_{s}$ ($|0\rangle_{s}$
and $|1\rangle_{s}$ denote that there are no photon and one photon, respectively) and a coherent probe beam in the state $|\alpha\rangle$
couple with a cross-Kerr nonlinearity medium, the systemic evolve as $\alpha|0\rangle_{s}|\alpha\rangle+\beta|1\rangle_{s}|\alpha e^{i\theta}\rangle$,
where $\theta=\chi t$ and $t$ is interaction time. The coherent probe beam picks up a phase shift directly proportional to the number of photons.
Hence, for the Eq.~(\ref{08}), one can find immediately that $|RLR\rangle$,
$|LRR\rangle$ and $|RRL\rangle$ cause the coherent beam $|\alpha\rangle$ to pick up a phase shift $\theta$, and $|LLL\rangle$
to pick up a phase shift $3\theta$. The different phase shifts can be distinguished by a general $X$ homodyne measurement. If we obtains
the state $|\alpha e^{i\theta}\rangle$, the three photons are in the maximally entangled state, with the probability
of 3/4. If we obtains the state $|\alpha e^{i3\theta}\rangle$, the three photons are in the state $|LLL\rangle$.

For the state $|LLL\rangle$, we let the second photon pass through a HWP, QWP and perform a CNOT gate on photons 1 and 2, with photon 2 act
as control qubit and photon 1 act as target qubit. Then the state $|LLL\rangle$ becomes $\frac{1}{\sqrt{2}}(|RLL\rangle+|LRL\rangle)$.
After that, we perform the operations in dashed box in Fig.~\ref{P3}. We can recover Eq.~(\ref{08}) again. So it still can be
used for conversion. At last, we convert the three photons to a $W$ state. The success probability of obtaining $W$ state
for the second round is $P=\frac{1}{4}\times\frac{3}{4}=\frac{3}{16}$. That is, by iterating the conversion process $n$ times, the total
success probability of this conversion is
\begin{eqnarray}\label{09}
P=\sum_{m=1}^{n}\left(\frac{1}{4}\right)^{m-1}\times\frac{3}{4},
\end{eqnarray}
It is obvious that the success probability increases with the increasing of the iteration time. When the iteration time equal to 4, the success probability can reach 99.6\%.

\subsection{Conversion of a four-photon GHZ state to a four-photon $W$ state}
\begin{figure}
\centering\scalebox{0.45}{\includegraphics{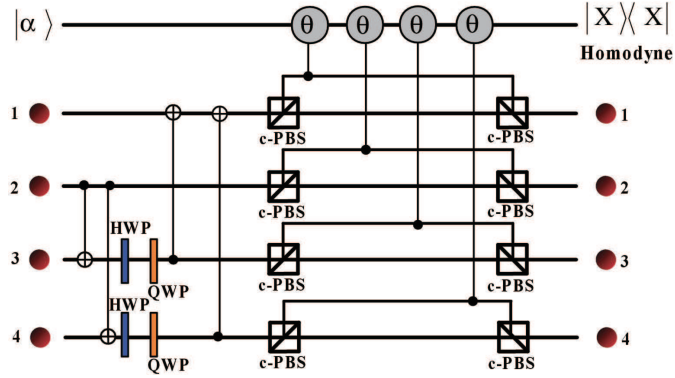}}
\caption{\label{P4}Schematic setup for converting a four-photon GHZ state to a four-photon $W$ state.}
\end{figure}
For a four-photon GHZ state convert to a four-photon $W$ state as shown Fig.~\ref{P4}. We assume the four photons are in the state
\begin{eqnarray}\label{10}
|\psi\rangle_{0}=\frac{1}{\sqrt{2}}(|RLRR\rangle+|LRLL\rangle)|\alpha\rangle,
\end{eqnarray}
After the four photons pass through the first two CNOTs, the systemic state becomes
\begin{eqnarray}\label{11}
|\psi\rangle_{1}=\frac{1}{\sqrt{2}}(|RLLL\rangle+|LRLL\rangle)|\alpha\rangle,
\end{eqnarray}
Then the photons pass through HWPs, QWPs and the second two CNOTs, the systemic state becomes
\begin{eqnarray}\label{12}
|\psi\rangle_{3}=\frac{1}{2\sqrt{2}}(|RRRL\rangle+|RRLR\rangle+|RLRR\rangle+|LRRR\rangle+|RLLL\rangle+|LRLL\rangle+|LLRL\rangle+|LLLR\rangle)|\alpha\rangle,
\end{eqnarray}
After the single photon pulse and a coherent probe beam in the state $|\alpha\rangle$ couple with a cross-Kerr nonlinearity medium,
the evolution of the whole system can be described as
\begin{eqnarray}\label{13}
|\psi\rangle_{3}&=&\frac{1}{2\sqrt{2}}(|RRRL\rangle+|RRLR\rangle+|RLRR\rangle+|LRRR\rangle)|\alpha e^{i\theta}\rangle \cr
&&+\frac{1}{2\sqrt{2}}(|RLLL\rangle+|LRLL\rangle+|LLRL\rangle+|LLLR\rangle)|\alpha e^{i3\theta}\rangle,
\end{eqnarray}
From the Eq.~(\ref{13}), one can see that whatever the measurement result is $|\alpha e^{i\theta}\rangle$ or $|\alpha e^{i3\theta}\rangle$, the
standard four-photon $W$ state can be obtained. Hence, the success probability for conversion of a four-photon GHZ state to a four-photon $W$ state
can reach unit.

\subsection{Conversion of a five-photon GHZ state to a five-photon $W$ state and $|D_{5}^{(2)}\rangle$ state}
\begin{figure}
\centering\scalebox{0.5}{\includegraphics{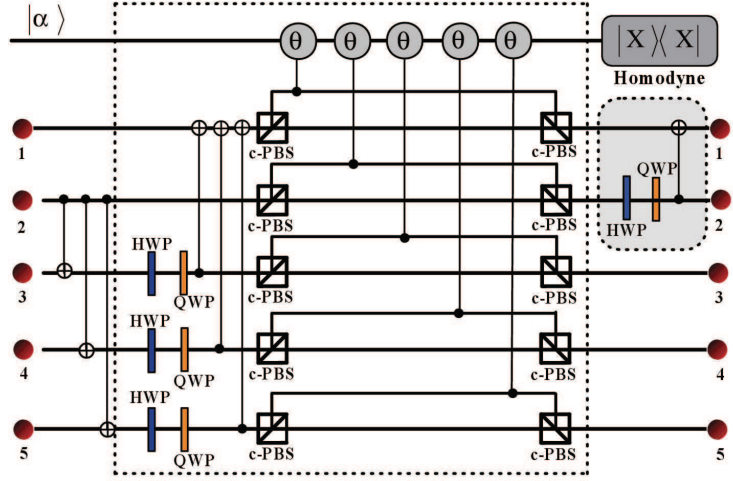}}
\caption{\label{P5}Schematic setup for converting a five-photon GHZ state to a five-photon $W$ state and $|D_{5}^{(2)}\rangle$ state.}
\end{figure}
For a five-photon GHZ state convert to a five-photon $W$ state as shown Fig.~\ref{P5}. We assume the five photons are in the state
\begin{eqnarray}\label{14}
|\psi\rangle_{0}=\frac{1}{\sqrt{2}}(|RLRRR\rangle+|LRLLL\rangle)|\alpha\rangle,
\end{eqnarray}
After the photons pass through the first three CNOTs, the systemic state becomes
\begin{eqnarray}\label{15}
|\psi\rangle_{1}=\frac{1}{\sqrt{2}}(|RLLLL\rangle+|LRLLL\rangle)|\alpha\rangle,
\end{eqnarray}
Then the photons pass through HWPs, QWPs and the second three CNOTs, the systemic state becomes
\begin{eqnarray}\label{16}
|\psi\rangle_{2}&=&\frac{1}{4}(|LRRRR\rangle+|RLRRR\rangle+|RRLRR\rangle+|RRRLR\rangle+|RRRRL\rangle+|LLLLL\rangle \cr
&&+|LLRRL\rangle+|LLRLR\rangle+|RLRLL\rangle+|LLLRR\rangle+|RLLRL\rangle+|RLLLR\rangle \cr
&&+|LRRLL\rangle+|LRLRL\rangle+|LRLLR\rangle+|RRLLL\rangle)|\alpha\rangle
\end{eqnarray}
Then the photon pulse and a coherent probe beam couple with a cross-Kerr nonlinearity medium,
the whole systemic state can be described as
\begin{eqnarray}\label{17}
|\psi\rangle_{3}&=&\frac{1}{4}\left(|LRRRR\rangle+|RLRRR\rangle+|RRLRR\rangle+|RRRLR\rangle+|RRRRL\rangle\right)|\alpha e^{i\theta}\rangle
+\frac{1}{4}|LLLLL\rangle|\alpha e^{i5\theta}\rangle\cr
&&+\frac{1}{4}(|LLRRL\rangle+|LLRLR\rangle+|RLRLL\rangle+|LLLRR\rangle+|RLLRL\rangle+|RLLLR\rangle \cr
&&+|LRRLL\rangle+|LRLRL\rangle+|LRLLR\rangle+|RRLLL\rangle)|\alpha e^{i3\theta}\rangle,
\end{eqnarray}

If the probe mode is in the state $|\alpha e^{i\theta}\rangle$, that is, the probe beam only picks up a phase shift $\theta$, the signal mode
will be projected to
\begin{eqnarray}\label{18}
|\varphi\rangle_{1}=\frac{1}{\sqrt{5}}\left(|LRRRR\rangle+|RLRRR\rangle+|RRLRR\rangle+|RRRLR\rangle+|RRRRL\rangle\right).
\end{eqnarray}
The conversion of a five-photon GHZ state to a $W$ state will succeed, which takes place with the probability 5/16.
However, if the probe mode is in the state $|\alpha e^{i3\theta}\rangle$, the signal mode will be projected to
\begin{eqnarray}\label{19}
|\varphi\rangle_{2}&=&\frac{1}{\sqrt{10}}(|LLRRL\rangle+|LLRLR\rangle+|RLRLL\rangle+|LLLRR\rangle+|RLLRL\rangle+|RLLLR\rangle \cr
&&+|LRRLL\rangle+|LRLRL\rangle+|LRLLR\rangle+|RRLLL\rangle)
\end{eqnarray}
Then it converts to another kind of entangled state, i.e., Dicke state $|D_{5}^{(2)}\rangle$.
The success probability of obtaining the $|D_{5}^{(2)}\rangle$ state is 5/8. If the probe mode is in the state $|\alpha e^{i5\theta}\rangle$,
the signal mode will be projected to $|LLLLL\rangle$. Here, we use the previous method, i.e., we let the second photon pass through a HWP, QWP
and perform a CNOT gate on photons 1 and 2. Then the state $|LLLLL\rangle$ becomes $\frac{1}{\sqrt{2}}(|RLLLL\rangle+|LRLLL\rangle)$.
After that, we perform the operations in dashed box in Fig.~\ref{P5}. We can obtain the Eq.~(\ref{17}) again. So it is still can be
used for conversion. At last, after $n$ times iteration, the total success probability of obtaining $W$ state is $P_{W}=\sum_{m=1}^{n}\left(\frac{1}{16}\right)^{m-1}\times\frac{5}{16}$ and the total success probability of obtaining $|D_{5}^{(2)}\rangle$
state is $P_{D}=\sum_{m=1}^{n}\left(\frac{1}{16}\right)^{m-1}\times\frac{10}{16}$. After many iterations, $P_{W}\simeq\frac{1}{3}$ and
$P_{D}\simeq\frac{2}{3}$.

\section{Analysis and discussion}
\begin{figure}
\begin{minipage}[ht]{0.49\linewidth}
\centering\includegraphics[width=\textwidth]{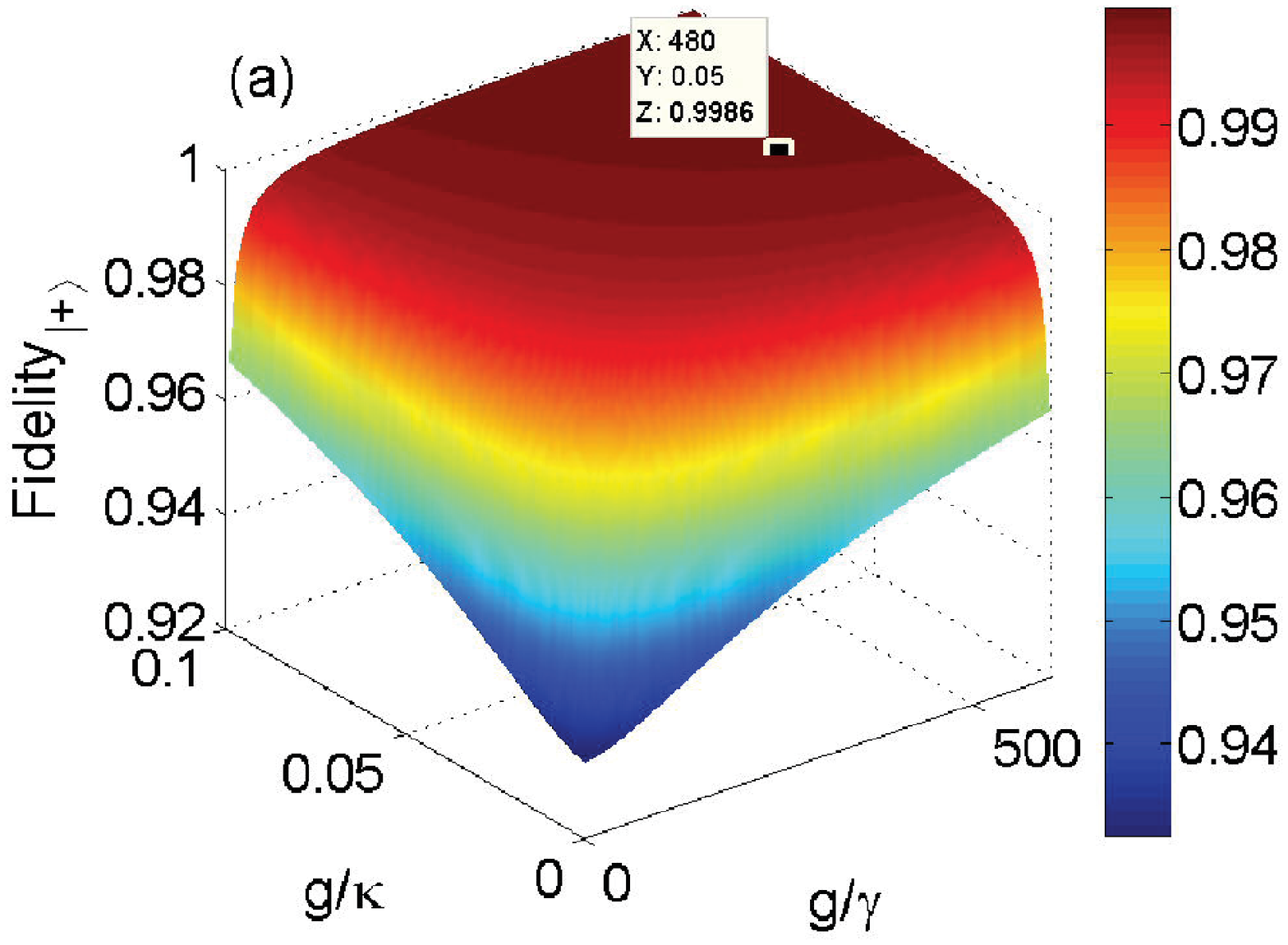}
\end{minipage}
\hfill
\begin{minipage}[ht]{0.49\linewidth}
\centering\includegraphics[width=\textwidth]{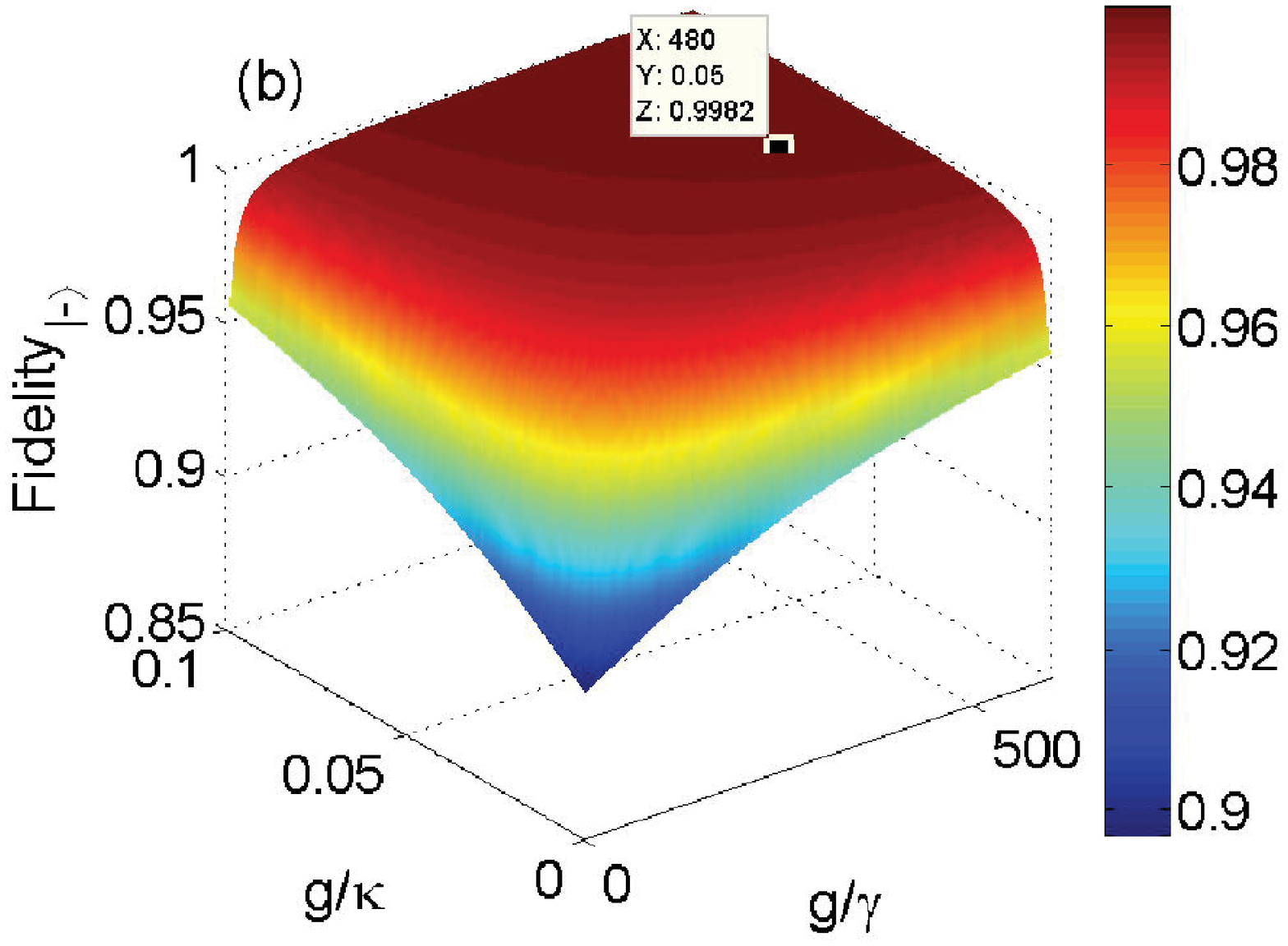}
\end{minipage}
\hfill
\begin{minipage}[ht]{0.49\linewidth}
\centering\includegraphics[width=\textwidth]{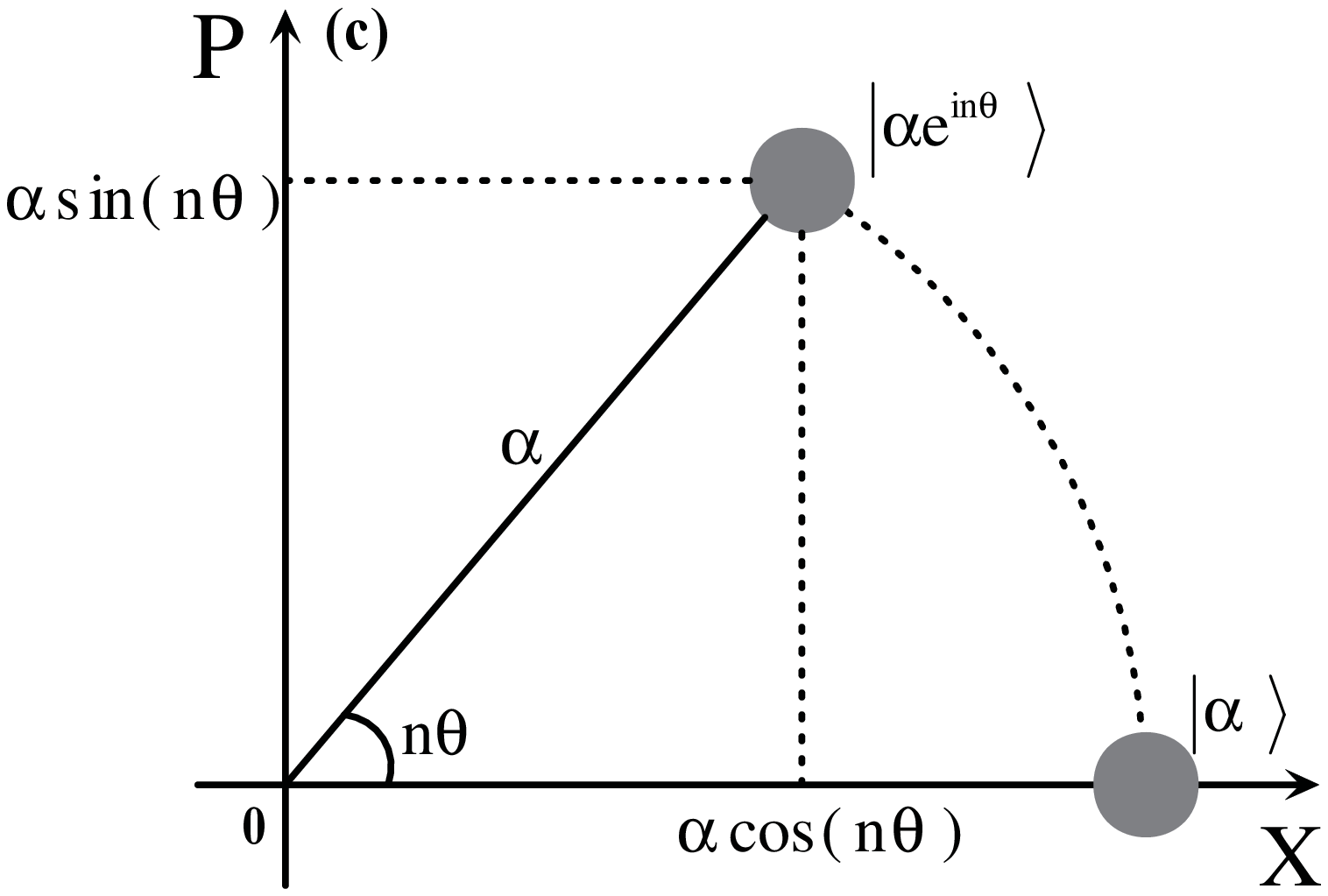}
\end{minipage}
\hfill
\begin{minipage}[ht]{0.49\linewidth}
\centering\includegraphics[width=\textwidth]{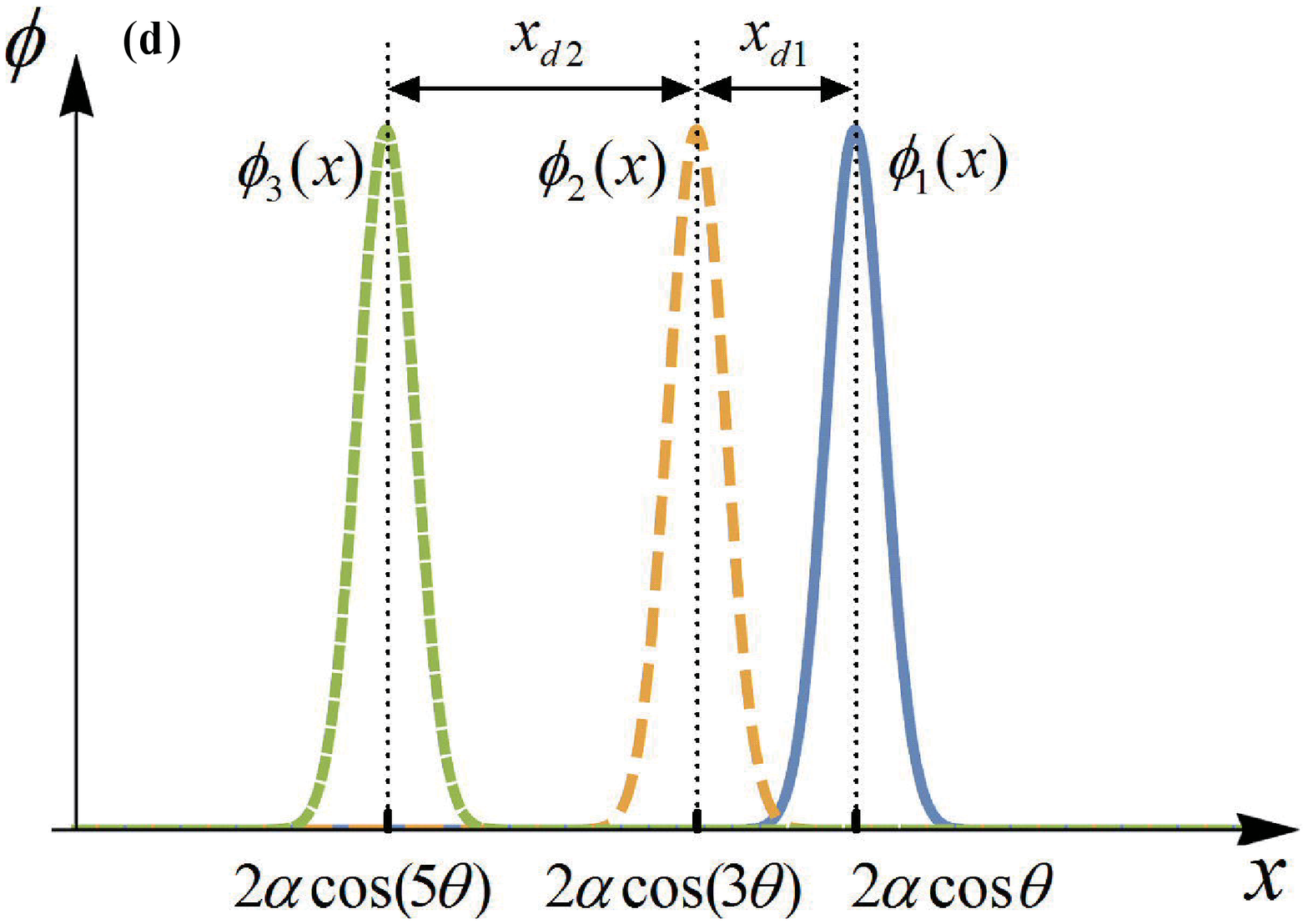}
\end{minipage}
\caption{\label{P6} (a)The fidelity of the CNOT gate versus the $g/\kappa$ and $g/\gamma$ corresponding to the measurement result of the N-V
center is $|+\rangle$. (b) The fidelity of the CNOT gate versus the $g/\kappa$ and $g/\gamma$ corresponding to the N-V center is $|-\rangle$.
(c)Schematic phase space illustration of the coherent state $|\alpha e^{in\theta}\rangle$. (d)Gaussian probability distribution for the result of the $X$ quadrature homodyne measurement. The blue solid curve, yellow dashed curve and the green dashed curve correspond to $\phi(x,\alpha\cos\theta)$, $\phi(x,\alpha\cos3\theta)$ and $\phi(x,\alpha\cos5\theta)$, respectively. $x_{d}$ is the distance between two peaks.}
\end{figure}

The schemes for conversion of entangled states are all constructed with the CNOT gate and cross-Kerr nonlinearity. Therefor the fidelity
of the CNOT gate and the error probability of the $X$ homodyne measurement are crucial to the schemes of the conversion.
Due to the imperfect coupling strength and leak rate will inevitably bring about slight influence
on the reflectance of output photons. The interaction of the incident photon pulse and N-V center in Eq.~(\ref{04})
should be rewritten as
\begin{eqnarray}\label{20}
&&|R\rangle|+\rangle\rightarrow-r_{0}(\omega_{p})|R\rangle|+\rangle,~~~~|R\rangle|-\rangle\rightarrow-r_{0}(\omega_{p})|R\rangle|-\rangle, \cr
&&|L\rangle|+\rangle\rightarrow-r_{0}(\omega_{p})|R\rangle|+\rangle,~~~~|L\rangle|-\rangle\rightarrow r(\omega_{p})|L\rangle|-\rangle.
\end{eqnarray}
The fidelity is defined as $|\langle \psi_{r}|\psi_{i}\rangle|^{2}$ to check the performance of the CNOT gate, in which $|\psi_{r}\rangle$
and $|\psi_{i}\rangle$ represent the final state in the realistic condition and the ideal condition, respectively. The Figs.~\ref{P6} (a) and ~\ref{P6}(b)
are the fidelities of the CNOT gate versus the $g/\kappa$ and $g/\gamma$ corresponding to the measurement result of the N-V center is $|+\rangle$
and $|-\rangle$ respectively, and show our schemes can be achieved with high fidelities. In \cite{046}, the parameters are chosen as follows: [$g,\kappa,\gamma_{total},\gamma_{ZPL}$]=[0.3,26,0.013,0.0004] GHz for a hybrid diamond-GaP microdisk system. Here, $\gamma_{ZPL}$ is the zero
phonon line (ZPL) emission spontaneous rate of an N-V center. By substituting these values, we have the fidelities for CNOT gate is 99.6\% and 99.5\%
corresponding to the measurement result is $|+\rangle$ and $|-\rangle$ respectively. Therefore, it has an effect on the fidelities of the conversion. For example, in the process of conversion of a three-photon GHZ state to a three-photon $W$ state, two CNOT gates need be performed, the fidelity can achieve 99.2\%, even the iteration time equal to four, the fidelity can also achieve 95.7\%. For conversion of a four-photon GHZ state to a four-photon $W$ state, we need not the iteration, so four CNOT gates need be performed, the fidelity can achieve 98.4\%. However, when the conversion of a five-photon GHZ state to a five-photon $W$ state and Dicke state is iterated four times, the fidelity can only achieve 89.7\%. 

In addition, in order to completely distinguish $|\alpha e^{i\theta}\rangle$, $|\alpha e^{i3\theta}\rangle$ and $|\alpha e^{i5\theta}\rangle$, we adopt homodyne measurement on the coordinate space of the coherent state. The wave function of the coherent state $|\alpha e^{in\theta}\rangle $ in the coordinate space
is $\phi(x,\alpha\cos(n\theta))=(2\pi)^{-\frac{1}{4}}\exp\left[-\frac{1}{4}(x-2\alpha\cos(n\theta))^{2}\right]$. Here, $\phi(x,\alpha\cos\theta)=\phi_{1}(x)$, $\phi(x,\alpha\cos3\theta)=\phi_{2}(x)$ and $\phi(x,\alpha\cos5\theta)=\phi_{3}(x)$, as shown in Figs.~\ref{P6} (c) and \ref{P6} (d), are Gaussian probability amplitudes. Therefore, the three Gaussian curves with three peaks located at $2\alpha\cos\theta$, $2\alpha\cos3\theta$ and $2\alpha\cos5\theta$. The error probability can be obtained with the same method as in \cite{047}. The small overlap between the two Gaussian curves amounts to the error probability, given by $P_{error}=\frac{1}{2}{\rm{erfc}}\left(x_{d}/2\sqrt{2}\right)$,
where $x_{d}$ is the distance between two peaks with $x_{d1}=2\alpha(\cos\theta-\cos3\theta)$ and $x_{d2}=2\alpha(\cos3\theta-\cos5\theta)$.
As a very promising method for generating the form of nonlinearity required, we consider a two-dimensional photonic crystal waveguide constructed
from diamond thin film with N-V centers fabricated in the center of the waveguide channel~\cite{048,049} could provide a sufficiently large nonlinearity.
For example, when the probe photon number $\alpha^{2}=1.3\times10^{4}$ and modest detunings, the cryogenic N-V-diamond system with $2\times10^{4}$
color centers can generate a phase shift of more than 0.1 rad per signal photon. Hence, the error probability for $X$ homodyne measurement in our
scheme is less than $10^{-5}$, which makes the scheme is feasible in the experiment with weak cross-Kerr nonlinearities.

In conclusion, we have proposed a scheme, based on single-photon input-output process and cross-Kerr nonlinearities, that allowed to locally convert
three-photon, four-photon and five-photon GHZ state to a $W$ state or Dicke state via LOCC. The total success probability can be improved by
iterating the conversion process for case of three-photon and five-photon. This method also can be used to convert a (N>5)-photon GHZ state into W state or Dicke state. The proposed schemes have unique characteristics compared with the existing
ones. The final analysis shows that our schemes are feasible and may have good performance in the current experimental conditions. We hope that our scheme could find some applications in the near future.

\section*{Acknowledgments}
We would like to thank Dr. Zhao Jin for valuable discussions.

\section*{Funding}
National Natural Science Foundation of China (NSFC) under Grants No. 11534002 and No. 61475033;
Fundamental Research Funds for the Central Universities under Grants No. 2412016KJ004.

\end{document}